\documentclass[aip,apl,reprint]{revtex4-1}

\usepackage{graphicx}
\usepackage{dcolumn}
\usepackage{bm}
\usepackage{multirow}

\begin{document}


\title{The Effect of Particle Strength on the Ballistic Resistance of Shear Thickening Fluids}


\author{Oren E. Petel}
\email[]{oren.petel@mail.mcgill.ca}
\affiliation{McGill University, Department of Mechanical Engineering, Montr\'{e}al, QC, Canada}
\author{Simon Ouellet}
\affiliation{Defence Research and Development Canada Valcartier, Qu\'{e}bec, QC, Canada}
\author{Jason Loiseau}
\affiliation{McGill University, Department of Mechanical Engineering, Montr\'{e}al, QC, Canada}
\author{Bradley J. Marr}
\affiliation{McGill University, Department of Mechanical Engineering, Montr\'{e}al, QC, Canada}
\author{David L. Frost}
\affiliation{McGill University, Department of Mechanical Engineering, Montr\'{e}al, QC, Canada}
\author{Andrew J. Higgins}
\affiliation{McGill University, Department of Mechanical Engineering, Montr\'{e}al, QC, Canada}



\begin{abstract}The response of shear thickening fluids (STFs) under ballistic impact has received considerable attention due to its field-responsive nature. While efforts have primarily focused on traditional ballistic fabrics impregnated with these fluids, the response of pure STFs to penetration has received limited attention. In the present study, the ballistic response of particle-based STFs is investigated and the effects of fluid density and particle strength on ballistic performance are isolated. It is shown that the loss of ballistic resistance in the STFs at higher impact velocities is governed by the material strength of the particles in suspension. The results illustrate the range of velocities over which these STFs may provide effective armor solutions.\end{abstract}



\maketitle


The integration of shear thickening fluids (STFs) in armor systems, a concept reported as early as Gates,\cite{Gates} has received considerable interest with the recent efforts to embed STFs within ballistic fabrics,\cite{Lee2003,Tan,Kalman,Park2} which has been shown to increase their ballistic performance; however, experimental evidence also suggests performance limitations of these hybrid armour systems. This loss of performance is evident when considering steel-core projectiles \cite{Gates, Park2} particularly when multiple layers and higher impact velocities are considered.\cite{Tan} Park et al.\cite{Park2} discussed preliminary experimental results in which a loss of effectiveness was seen against steel projectiles (FSPs) above an impact velocity of 300~m/s, the same velocity range investigated by Tan et al.~\cite{Tan} The coupled nature of the fluid-fabric interactions make it difficult to ascertain whether this behavior is due to a loss of performance within the STF itself or a transition in the dominant failure mode within the fibers, rendering the presence of the STF inconsequential to the ballistic response. In the present study, we investigate the ballistic penetration of several STFs, particularly focusing on the role of particle strength in determining the ballistic response of STFs through variations of the particle material and volume fraction in the suspensions.

\begin{figure}[Ht]
 \includegraphics{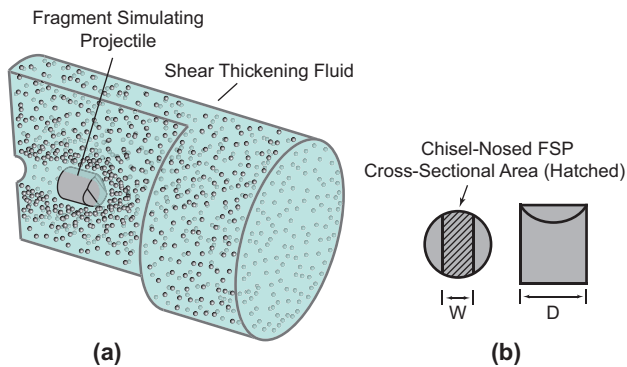}
 \caption{Schematic of (a)~a broken-out section view of the FSP penetrating the STF. (b)~Top and side views of the FSP.
\label{fig1}}
\end{figure}

STFs are field-responsive fluids which can undergo a sudden fluid-solid transition under certain stimuli. STFs have been extensively characterized using low-stress dynamic techniques,\cite{Hoffman,Barnes,Lim2} in which liquids are considered incompressible. These conditions are not directly relevant to the dynamic high-stress environment of a ballistic impact, where compressibility effects dominate material responses.\cite{Field} Lee and Kim~\cite{LeeKim} estimated the stagnation pressure at the nose of a steel projectile impacting a STF-impregnated fabric to be on the order of several gigapascals, stresses at which compressibility effects must be considered. Under ballistic conditions, in addition to traditional shear thickening mechanisms, a compression-induced clustering of particles should be expected as the liquid density and particle volume fraction increase under high pressures,\cite{PetelJAP,PetelPRE,PetelAPS1,PetelAPS2} resulting in extensive particle force chains forming around the projectile (Fig.~\ref{fig1}a and \ref{fig1}b) as the solid phase volume fraction in the STF can increase by as much as 10$\%$ due to the impact pressures. The \textit{in situ} formation of force chains implies that the ballistic response of various STFs should be directly related to the strength of the suspended particles.

We investigated this hypothesis by studying the ballistic penetration of several STFs, a dilute suspension (non-shear-thickening), and neat ethylene glycol (the suspending medium for the mixtures considered). The suspended particulate phase of the STFs was varied in a manner that interrogated the influence of the material strength of the particles on penetration resistance. The experimental ballistic penetration results are presented in reference to an inertially-based penetration model to highlight the strength limitations of the various STFs.

\begin{figure}
 \includegraphics{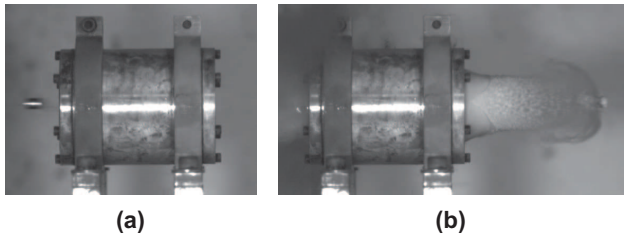}
 \caption{Photographs of the FSP (a) entering and (b) exiting the test capsule. \label{fig2}}
\end{figure}

The penetration resistance was investigated by measuring the ability of the fluids to decelerate a 17 grain (1.1~g) chisel-nosed mild steel NATO-standard fragment simulating projectile (FSP), shown schematically in Fig.~\ref{fig1}c, in a configuration similar to that used by Nam et al.\cite{Nam} The samples were tested in a cylindrical aluminum capsule with an internal diameter of 38~mm and a length of 64~mm. Mylar diaphragms (0.1~mm thick), which were found to have a negligible influence on the experimental results, were used to confine the fluid samples in the capsules during the experiments. The target capsule was positioned close to the end of the smooth-bore gas-gun barrel in order to minimize the projectile yaw at the impact face. Experiments were conducted with a range of incident FSP velocities of 200 to 700~m/s. The incident and residual velocities of the FSP were measured with a Photron SA5 high-speed camera at 20,000~fps. A set of images taken of the incident and exiting projectile are shown in Fig.~\ref{fig2}a and \ref{fig2}b respectively.

The investigation involved several mixtures, the proportions of which are given in Table~\ref{Tab1}. Particle settling was not a concern as the capsules were filled 5-10 minutes prior to the experiments and vortex mixers were used to ensure sufficient dispersion of the particles. The components of the various mixtures included liquid ethylene glycol (EG), silica (Fiber Optic Center, monodisperse spheres, $d$~=~1~$\mu$m), $\alpha$-silicon carbide (Washington Mills, irregular morphology, $d_{\mathrm{mean}}$~=~5~$\mu$m), and cornstarch (Fleischmann, $d_{\mathrm{mean}}$~=~10~$\mu$m). The three types of solid particles have drastically different material properties (Table~\ref{Tab2}). In order of increasing strength, the materials are cornstarch, silica, and silicon carbide. 

It should be noted that the silica particles used in the present study had a nano-porous structure consisting of voids which were inaccessible to the ethylene glycol. This void fraction resulted in a wetted bulk particle density of 1.85~g/cm$^{3}$, effectively containing a 16$\%$ gas-filled void fraction. The bulk particle density of the silica particles used in the present study is consistent with the wetted bulk density of silica particles used in previous ballistic experiments involving silica-based STFs.\cite{Lee2003} This significant gas-filled void fraction would have an adverse affect on the strength of the silica particles in comparison to the bulk-material values listed in Table~\ref{Tab2}.

\begin{table}
	\centering
	\caption{\label{Tab1}Summary of the mixture compositions investigated in the present study. All mixtures contained ethylene glycol as the suspending medium.}
	\begin{ruledtabular}
	\begin{tabular}{lllll}
	\multirow{2}{*}{\textbf{\small{Sample}}}&\textbf{\small{Solid}}&\textbf{\small{Solid Volume}} &\textbf{\small{Density}}&\multirow{2}{*}{\textbf{\small{STF}}}\\
	&\textbf{\small{Component(s)}}&\textbf{\small{Fraction ($\%$)}}&\textbf{\small{(g/cm$^{3}$)}}&\\
	\small{EG}&\small{n/a}&\small{n/a}&\small{1.11}&\small{No}\\
	\small{54~CS}&\small{Cornstarch}&\small{54.0}&\small{1.35}&\small{Yes}\\
	\small{21~SiC}&\small{Silicon Carbide}&\small{21.5}&\small{1.57}&\small{No}\\
	\small{61~SiO$_{2}$}&\small{Silicon Dioxide}&\small{61.5}&\small{1.57}&\small{Yes}\\\\
	\multirow{2}{*}{\small{61~Mix}}&\small{Silicon Dioxide}&\small{47.6}&\multirow{2}{*}{\small{1.76}}&\multirow{2}{*}{\small{Yes}}\\
	&\small{Silicon Carbide}&\small{13.9}&&\\
	\end{tabular}
	\end{ruledtabular}
\end{table}

Of the mixtures that are summarized in Table~\ref{Tab1}, three of the mixtures investigated exhibit shear thickening behaviors: 54~CS, 61~SiO$_{2}$, and 61~Mix. The 21~SiC mixture, a dilute suspension that did not exhibit shear thickening behavior, was used for a density-matched comparison to the 61~SiO$_{2}$ mixture. The velocity decrement of the FSP, the difference between the incident and residual velocities of the FSP penetrating the fluids, was used as the basis of comparison between test mixtures. This measure provides a means of evaluating the competition between the mixture density and the material strength of the suspended solid phase as the dominant factor in the ballistic resistance of the various fluids. 

\begin{table}
	\centering
	\caption{\label{Tab2}Summary of the bulk-material properties for the solid materials used in the present study.}
	\begin{ruledtabular}
	\begin{tabular}{llll}
	\multirow{3}{*}{\textbf{\small{Material}}}&\textbf{\small{Density}}&\textbf{\small{Young's}}&\textbf{\small{Hardness}}\\
	&\textbf{\small{(g/cm$^{3}$)}}&\textbf{\small{Modulus}}&\textbf{\small{(GPa)}}\\
	&&\textbf{\small{(GPa)}}&\\
	\small{Cornstarch}&\small{1.55}&\small{0.053}\footnote{Agbisit et al.\cite{Agbisit}}&\small{n/a}\\
	\small{Silicon Dioxide}&\small{2.20}&\small{69.3}\footnote{Pharr\cite{Pharr}}&\small{8.3}$^{\mathrm{b}}$\\
	\small{Silicon Carbide}&\small{3.22}&\small{454.7}$^{\mathrm{b}}$&\small{30.8}$^{\mathrm{b}}$\\
\end{tabular}
	\end{ruledtabular}
\end{table}

A simple analytical penetration model can be used to predict an inertially-dominated penetration behavior in targets of various densities. The assumptions inherent in this model are: (\textit{i}) the projectile drives a plug through the target with a cross-sectional area equal to that of the chisel nose (Fig.~\ref{fig1}c), (\textit{ii}) the target material has no material strength (the hydrodynamic limit), and (\textit{iii}) the impact and penetration process is perfectly plastic, resulting in identical final velocities of the projectile and plug. The assumption concerning the cross-sectional area of the plug accounts for the divergence of material around the projectile tip. The model can therefore be adequately described by conserving momentum through the equation,\begin{equation}
\rho_{p} L_{p} \cdot V_{i} = \left(\rho_{p} L_{p} + \rho_{t} L_{t}A_{r}\right) \cdot V_{f}
\label{eq1}
\end{equation} where $L$ is the length, $\rho$ is the density, $A_{r}$ is the area ratio of the chisel nose to the projected area of the FSP, $V_{i}$ is the incident projectile velocity, $V_{f}$ is the residual projectile velocity, and the subscripts $p$ and $t$ refer to the projectile and target respectively. The area ratio can be determined from geometric considerations such that,\begin{eqnarray}
&A_{r}=1 -\frac{1}{\pi}\cdot\left(\theta-\mathrm{sin}\theta\right)\\
&\theta=2\mathrm{cos}^{-1}\left(\frac{W}{D}\right)
\label{eq2}
\end{eqnarray} where $D$ is the diameter and $W$ is the width of the chisel nose, 5.38~mm and 2.54~mm, respectively, for the FSP. 

\begin{figure}
 \includegraphics{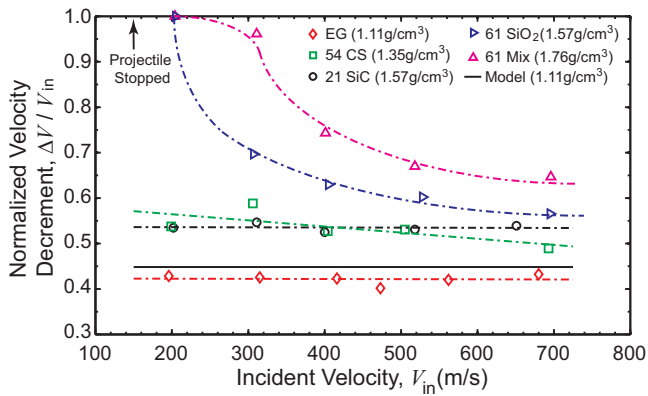}
 \caption{A comparison of the change in FSP velocity normalized by the incident velocity for various mixtures. \label{fig3}}
\end{figure}

This penetration model predicts that for a given target density and capsule size, the velocity decrement through the sample, normalized by the incident projectile velocity, is independent of the incident FSP velocity (Fig.~\ref{fig3}). The invariance of this parameter provides a means of investigating the effect of dynamic material strength on ballistic penetration through a comparison of the experimental results to the inertially-based penetration predictions.

The result of the model, calculated for a target with the density of ethylene glycol (EG), is represented in Fig.~\ref{fig3} by a solid line. Experimentally, three of the samples tested in the present study (EG, 21~SiC, and 54~CS) were found to follow the trend predicted by the model, despite the fact that 54~CS is an STF. The two mixtures which significantly deviated from this trend were the 61~SiO$_{2}$ and 61~Mix, both STFs containing particles with considerable material strength. 

The 61~SiO$_{2}$ and 61~Mix mixtures showed similar deviations from the behavior predicted by the model.  At lower velocities, both mixtures were effective at stopping or decelerating the projectile. At higher velocities, their response asymptoted toward an inertially-dominated deceleration, independent of further increases in the incident projectile velocity. Therefore, the results point to a transition in penetration response in these mixtures, where the low velocity penetration is dominated by the dynamic material strength of the mixtures and the higher velocity penetration is governed by the mixture densities. 

\begin{figure}
 \includegraphics{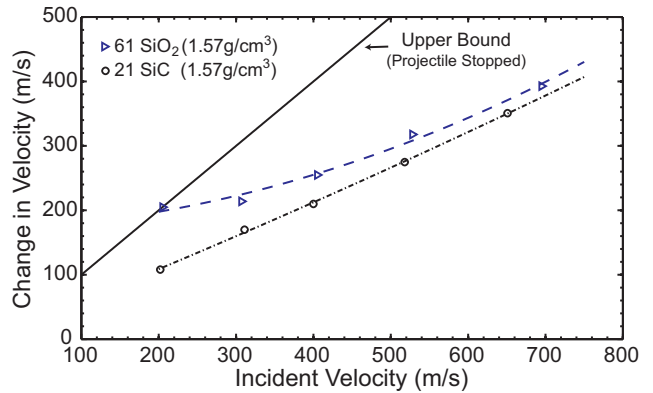}
 \caption{A comparison of the response of two mixtures with the same bulk density, 21~SiC and 61~SiO$_{2}$, which isolates the shear thickening contribution to penetration resistance. \label{fig4}}
\end{figure}

This transition is particularly evident when comparing two mixtures with the same density, 61~SiO$_{2}$ and 21~SiC, the former being a STF (Fig.~\ref{fig4}). At low impact velocities, the ballistic results suggest that the onset of a dynamic material strength in 61~SiO$_{2}$ governs its resistance to penetration. However, at higher impact velocities, this effect is less important and the bulk mixture density dominates the response, evidenced by the convergence of the results for the two mixtures.

Comparing the results for the three STFs on the basis of their transition to a inertially-dominated behavior in Fig.~\ref{fig3}, the effect of the particle strength is clearly evident. While the mixtures containing ceramic particles (61~SiO$_{2}$ and 61~Mix) have a noticeable material strength response at lower impact velocities, no such effect is seen in the results with 54~CS targets. Due to the low yield strength of cornstarch in comparison to the two ceramics, the 54~CS mixture exhibits a response similar to the trend of the penetration model and the 21~SiC and EG experiments.

Considering the results of the 61~SiO$_{2}$ and 61~Mix (Fig.~\ref{fig3}), the particle strength effect on ballistic penetration is most pronounced. Both of these mixtures have identical solid phase volume fractions, the only difference being the proportion of silica and silicon carbide in the two mixtures (Table~\ref{Tab1}). The addition of a small volume of silicon carbide extended the incident velocity range over which the material strength dominated the ballistic response of the STF. The higher yield strength of the silicon carbide particles in comparison to that of porous silica led to the increased ballistic resistance of the 61~Mix STF. This increased resistance resulted in delaying the transition from material strength to inertially-dominated penetration resistance to a higher incident projectile velocity. It should be noted that the silica-based STF (61~SiO$_{2}$), demonstrated a loss of performance at approximately 300~m/s. This critical impact velocity is consistent with the loss of performance observed experimentally in STF-embedded fabrics,\cite{Tan,Park2} suggesting a relationship between this behavior and the observed loss of dynamic strength in the bulk STFs. However, it is understood that STF-impregnated fabrics are complex systems and the ballistic performance of the system is highly dependent on the fluid-fiber interactions.
 
As each STF mixture has a liquid and solid phase, the possibility that failure within either of these phases can lead to the observed loss of strength in the impacted fluids should be considered. In previous rheological studies of STFs, failure of shear thickened fluids in the form of cracking has been observed. It has been suggested that this behavior is the result of cavitation (tensile failure) of the liquid suspending medium.\cite{Franks} This observation is consistent with results obtained in saturated sand, where the mechanism leading to a loss of strength is highly dependent on the pore pressure of the fluid.\cite{Bless} When strained at low confining pressures, the dilatancy of the system can result in cavitation within the fluid.\cite{Bless,Witman} However, under high-strain-rate loading and high confining pressures, grain crushing influences the dynamic behavior of dense saturated sand.\cite{Bless,LeeSand} In the case of a ballistic impact, as the projectile enters the fluid, a shock wave is driven ahead of the projectile, resulting in a high-pressure region acting on the projectile face. In a ballistic impact, the conditions for cavitation at the projectile tip are not met due to the high pressure, however cavitation will often occur in the lateral and wake regions of the projectile path.\cite{Cavity1} Were the loss of strength observed in the present study primarily governed by cavitation, altering the solid phase material in the mixtures would have had a minimal influence on the critical incident velocity since the same liquid, ethylene glycol, was used in all mixtures. With these considerations, it is more probable that the loss of penetration resistance was related to the response of the solid phase of the mixtures. 

This picture is consistent with previous studies comparing wetted (STF) and dry particles integrated into ballistic fabrics, which found that both impregnation techniques were comparably effective at increasing the ballistic limit of the fabrics.\cite{Gates,Kalman} It has also been demonstrated the ballistic performance of impregnated Kevlars correlated to the material strength of the embedded particles, whereby introducing silica particles resulted in a higher ballistic limit than introducing softer polymethalmethacrylate particles.\cite{Kalman}

The development of stress-bearing disordered structures has been shown to occur in dense suspensions under either shear\cite{Hoffman,Barnes} or high-speed uniaxial compression loading.\cite{PetelJAP,PetelAPS1,PetelAPS2,PetelPRE} Experimental measurements of the dynamic shear strength of dense silicon carbide-based suspensions were found to be on the order of 0.5~GPa,\cite{PetelJAP} a significant deviation from a hydrodynamic stress state. A dynamic shear strength within the suspension is evidence of interparticle force chains influencing the bulk stress state within the fluids, as stress fluctuations along force chains can deviate from the mean stress by an order of magnitude.\cite{Behringer} For the higher velocity impacts, where stresses at the impact face can reach several gigapascals, the stress fluctuations at these interparticle contact points could be of sufficient strength to locally crush or deform the particles considered in the present study. Particle deformation in a wetted granular system loaded at high strain rates is consistent with previous observations.\cite{Bless,LeeSand,Yoshinaka} Preliminary data have been obtained by the present authors in plate impact tests with the 61~SiO$_{2}$ mixture that show evidence of particle crushing in the recovered samples following the transmission of a 1~GPa shock wave through the STF. Further testing investigating this effect is presently underway.

If we consider that the dynamic response of the suspensions involves mesostructural reorganization and the formation of clustered particles, the strength of the suspended particles becomes an important parameter in the dynamic response of the STF. The formation of force chains would limit the ability of the fluid in the STFs to flow until the force chains are destroyed, a process that would be dependent on the strength of the particles in the chains. Under low velocity impacts, where particle crushing is unlikely, force chains can be disrupted through rotational and translational mechanisms. At higher impact velocities, if the dynamic strength of the particles is exceeded, force chains can be disrupted through localized particle deformation (fracture or crushing), which would result in a loss of strength in the stiffened mixture. Once the strength of the materials forming the clusters within the STF is exceeded, the material response is inertially dominated.

The authors thank Jacques Blais of DRDC Valcartier for his assistance in conducting the experiments. This work was conducted with the support of Hamid Bennadi of Stedfast Ltd., NSERC, and the Department of National Defence under project DNDPJ 385687-09.


\end{document}